\documentclass[11pt]{article} 
\usepackage{cite}
\usepackage{amsmath,amsfonts}
\usepackage{bm}
\usepackage{mathrsfs}
\usepackage{latexsym}
\usepackage{amssymb}
\usepackage[breaklinks=true,linktocpage=true]{hyperref}
\hypersetup{colorlinks=true, allcolors=blue}
\usepackage{graphicx}

\usepackage{slashed}
\usepackage{mathbbol}

\usepackage{color}
\usepackage{comment}
\usepackage{physics}
\usepackage{float}
\usepackage{siunitx}

\begin{document}
\pagestyle{empty}
\vspace{-4cm}
\begin{center}
    \hfill YITP-25-134 \\
\end{center}

\vspace{2cm}

\begin{center}

{\bf\LARGE
Unitarity test of lepton mixing
via energy dependence of 
neutrino oscillation
}
\\

\vspace*{1.5cm}
{\large 
Ryuichiro Kitano$^{1}$, 
Joe Sato$^{2}$, and 
Sho Sugama$^{2}$
} \\
\vspace*{0.5cm}

{\it 
$^1$Yukawa Institute for Theoretical Physics, Kyoto University,
Kyoto 606-8502, Japan\\
$^2$Department of Physics, Facility of Engineering Science, Yokohama National University, Yokohama 240-8501, Japan 
}

\end{center}

\vspace*{1.0cm}

\begin{abstract}
{\normalsize
We study the method to test the unitarity of the lepton mixing matrix by using only the long baseline neutrino oscillation experiments, 
such as the combination of the T2HK experiment and the one with the $\nu_e$ beam from a future neutrino factory at J-PARC.
Without a specific parametrization, one can directly extract the elements of the lepton mixing matrix
by observing the energy dependence of the oscillation probabilities.
A non-trivial test of the unitarity under the three-generation assumption can thus be made possible by examining the orthogonality
in a similar manner to the unitarity triangle in the quark sector.
As the first trial, 
we perform the analysis based on the simplified situation where the matter effects in the neutrino oscillation can
be neglected.
Under this simplified analysis, we demonstrate that a violation of unitarity in the $3\times3$ lepton mixing matrix can be observed, using a four-generation model as a concrete example of a non-unitary three-generation scenario.
The statistically most significant measurement can be
provided by the energy dependences of
the combination of
the CP conjugate modes, $\nu_\mu \to \nu_e$ and $\bar \nu_\mu \to \bar \nu_e$, at
T2HK and, independently, by the T conjugate modes, 
$\nu_\mu \to \nu_e$ and $\nu_e \to \nu_\mu$, with the latter
measured at the
neutrino factory experiments.
}
\end{abstract} 

\newpage
\baselineskip=18pt
\setcounter{page}{2}
\pagestyle{plain}

\setcounter{footnote}{0}

\section{Introduction}
\label{sec1}
It has been established that neutrinos undergo oscillation phenomena caused by mixing among generations. Various measurements, starting from observations of solar and atmospheric neutrino oscillations~\cite{KamLAND:2010fvi,SNO:2011hxd,Vinyoles:2016djt,KamLAND:2013rgu,Cleveland:1998nv,GNO:2005bds,SAGE:2009eeu,Bellini:2011rx,Super-Kamiokande:2016yck,IceCube:2017lak,IceCube:2019dqi,Super-Kamiokande:2019gzr}, have determined parameters such as mixing angles and mass differences, with some remaining discrete ambiguities.

Despite the recent dramatic progress in measuring oscillation parameters, we can still say that studies exploring the nature of lepton mixing have only just begun. In addition to measuring the leptonic CP phase, it is important to start over-constraining the parameter space, similar to the quark sector, where the underlying theory has been pinned down to the $3 \times 3$ Cabibbo-Kobayashi-Maskawa (CKM)~\cite{Cabibbo:1963yz,Kobayashi:1973fv} paradigm.

The most fundamental aspects to examine would be the CPT invariance and unitarity in the leptonic sector. In the quark sector, the unitarity of the CKM matrix is visualized as the unitarity triangle~\cite{UTfit:2006vpt,Hocker:2001xe,HFLAV:2016hnz,Wolfenstein:1983yz,ParticleDataGroup:2018ovx,Charles:2004jd,Buras:1994ec}, which represents the orthogonality of any two column vectors in the CKM matrix. If the CKM matrix is unitary, the vertices obtained from various experiments should converge at a point in the complex plane. The area of the triangle is proportional to the Jarlskog invariant, which represents CP violation.

In the lepton sector, past and ongoing neutrino oscillation experiments~\cite{KamLAND:2010fvi,DayaBay:2018yms,SNO:2011hxd,Vinyoles:2016djt,NOvA:2019cyt,OPERA:2018nar,T2K:2018rhz,T2K:2019bcf,KamLAND:2013rgu,Cleveland:1998nv,GNO:2005bds,SAGE:2009eeu,Bellini:2011rx,Super-Kamiokande:2016yck,IceCube:2017lak,IceCube:2019dqi,Super-Kamiokande:2019gzr} have been measuring the parameters of the PMNS matrix~\cite{Pontecorvo:1967fh,Maki:1962mu}, assuming it to be a $3 \times 3$ unitary matrix. Future experiments~\cite{Hyper-Kamiokande:2018ofw,DUNE:2015lol,JUNO:2015zny,IceCube-Gen2:2020qha} are expected to further improve the precision of these measurements. Thus, precision tests of the unitarity of the leptonic mixing matrix are becoming feasible~\cite{Antusch:2006vwa,Qian:2013ora,Escrihuela:2015wra,Parke:2015goa}, and the construction of the leptonic unitarity triangle is also possible~\cite{Sato:2000wv,Farzan:2002ct,He:2013rba,Gonzalez-Garcia:2014bfa,He:2016dco,Esteban:2016qun,Ellis:2020ehi,Ellis:2020hus}.

In general, if there are $3+M$ generations of neutrinos, the lepton mixing matrix is a $(3+M) \times (3+M)$ unitary matrix. The $3 \times 3$ submatrix part is, of course, not unitary in general~\cite{Wyler:1982dd,Langacker:1988up,Hewett:1988xc,Buchmuller:1991tu,Ingelman:1993ve,Nardi:1993ag,Chang:1994hz,Tommasini:1995ii,Loinaz:2003gc}. Standard oscillation analyses assume the matrix is $3 \times 3$ and unitary, with three angles and a phase to parameterize the matrix. Fitting the data with this assumption does not test unitarity.

To perform the test, one should relax the assumption by increasing the number of parameters and check if the unitarity condition is satisfied. In principle, measurements of the energy dependence of oscillation probabilities could be useful, as different combinations of the elements of the lepton mixing matrix can be extracted without a specific parametrization~\cite{Sato:2000wv}.

In this paper, we explore such a method for testing the unitarity of the lepton mixing matrix independent of its parametrization. We demonstrate an analysis assuming a situation where $\nu_\mu$, $\bar{\nu}_\mu$, and $\nu_e$ beams are available. This assumption reflects the beam availability in future experiments such as T2HK~\cite{Hyper-Kamiokande:2018ofw} and a neutrino factory with a $\nu_e$ beam at J-PARC~\cite{Kitano:2024kdv,Hamada:2022mua}. For simplicity, we perform the analysis using vacuum neutrino oscillations because the baseline is relatively short. Another reason for this simplification is that, in the case of a neutrino factory, T-violation can also be observed. Since T-violation is less affected by matter effects, a vacuum-based demonstration combining $\nu_e \to \nu_\mu$ oscillations may capture the essential features of the full analysis.

This paper is organized as follows. In Section~\ref{sec2}, we review the lepton mixing matrix and neutrino oscillations and introduce the new criterion to test unitarity. In Section~\ref{sec3}, we conduct a statistical analysis to test the unitarity of the lepton mixing matrix. Finally, Section~\ref{sec4} summarizes our conclusions.

\section{The Leptonic Mixing Matrix and Neutrino Oscillations}
\label{sec2}
The leptonic mixing matrix describes the mixing between flavor eigenstates and mass eigenstates. In general, when $N$ generations of neutrinos mix with $M$ flavors, the leptonic mixing matrix $V$ is an $M\times N$ matrix.
By construction the $3 \times 3$ part of the matrix $V$ is not necessarily
unitary. Below we construct a parameter $\xi$ that vanishes
when the $3 \times 3$ part is unitary while it can be
measured by looking at the energy dependence of the neutrino
oscillation probabilities in vaccum.
We design the $\xi$ parameter so that the measurement at $\xi = 0$ provides a {\it sufficient} condition for unitarity.
In general, the lepton mixing
matrix is given by
\begin{align}
    \mqty(\nu_e \\ \nu_\mu \\ \nu_\tau \\ \vdots) 
    &= V \mqty(\nu_1 \\ \nu_2 \\ \nu_3 \\ \vdots) 
    \\
    &=\mqty(V_{e1} & V_{e2} & V_{e3} & \cdots 
           \\
           V_{\mu1} & V_{\mu2} & V_{\mu3} & \cdots
           \\
           V_{\tau1} & V_{\tau2} & V_{\tau3} & \cdots
           \\
           \vdots & \vdots & \vdots & \ddots) 
           \mqty(\nu_1 \\ \nu_2 \\ \nu_3 \\ \vdots)\ .
\end{align} 
If there are only three generations and three flavors, $V=U_{\rm PMNS}$. In the following, we denote the $3 \times 3$ $U_{\rm PMNS}$ matrix simply as $U$ while $V$ is meant to be a general $M \times N$ mixing matrix. This mixing matrix leads to neutrino oscillations.
In the standard
$3 \times 3$ unitary case,
there are 9 unitarity conditions from $UU^{\dag}=U^{\dag}U=\vb{1}$.
\begin{align}
    \sum_j U_{\alpha j}U_{\beta j}^{*} &= \delta_{\alpha \beta},
    \label{condition1}
    \\
    \sum_\alpha U_{\alpha j} U_{\alpha k}^{*} &= \delta_{jk}\ . 
    \label{condition2}
\end{align}
Since the matrix elements $U_{\alpha j}$ are complex, the conditions in Eq.~\eqref{condition1},~\eqref{condition2} correspond to constructing triangles in complex planes.

The neutrino oscillations in vacuum are described by the following equation,
\begin{align}
    i\dv{}{t}\mqty(\nu_e(\bar{\nu}_e)\\ \nu_\mu(\bar{\nu}_\mu) \\ \nu_\tau(\bar{\nu}_\tau) \\ \vdots) 
    &=V^{(*)} \mathrm{diag}(E_1, E_2, E_3,\ \ldots)V^{\dag(T)}
    \mqty(\nu_e(\bar{\nu}_e)\\ \nu_\mu(\bar{\nu}_\mu) \\ \nu_\tau(\bar{\nu}_\tau) \\ \vdots)\ .
    \label{eq:osc}
\end{align}
The oscillation probability for $\nu_\alpha(\bar{\nu}_\alpha) \to \nu_\beta(\bar{\nu}_\beta)$ are thus given by
\begin{align}
    P^{\nu_{\alpha\to\beta}}(P^{\bar{\nu}_{\alpha\to\beta}}) = \delta_{\alpha\beta}
    & -4\sum_{j>k}\mathrm{Re}\left[V_{\alpha j} V^{*}_{\beta j} V^{*}_{\alpha k} V_{\beta k}\right]
    \sin^2\left(\frac{\mathit{\Delta} E_{jk}L}{2}\right)\notag
    \\
    & \mp2\sum_{j>k}\mathrm{Im}\left[V_{\alpha j} V^{*}_{\beta j} V^{*}_{\alpha k} V_{\beta k}\right]\sin\left(\mathit{\Delta} E_{jk}L\right)\ .
    \label{eq:prob}
\end{align}
Here and in the following, we denote $\nu_\alpha \to \nu_\beta$ and $\bar{\nu}_\alpha \to \bar{\nu}_\beta$ by $\nu_{\alpha\to\beta}$ and $\bar{\nu}_{\alpha\to\beta}$.
The unitarity of $V$ is assumed in the formula.
It is a general formula up to here, but below we make assumptions which are not 
necessarily true in general models. Such treatment is justified since we are interested in deriving 
a necessary condition for unitarity.

\begin{figure}[H]
    \centering
    \includegraphics[width=10cm]{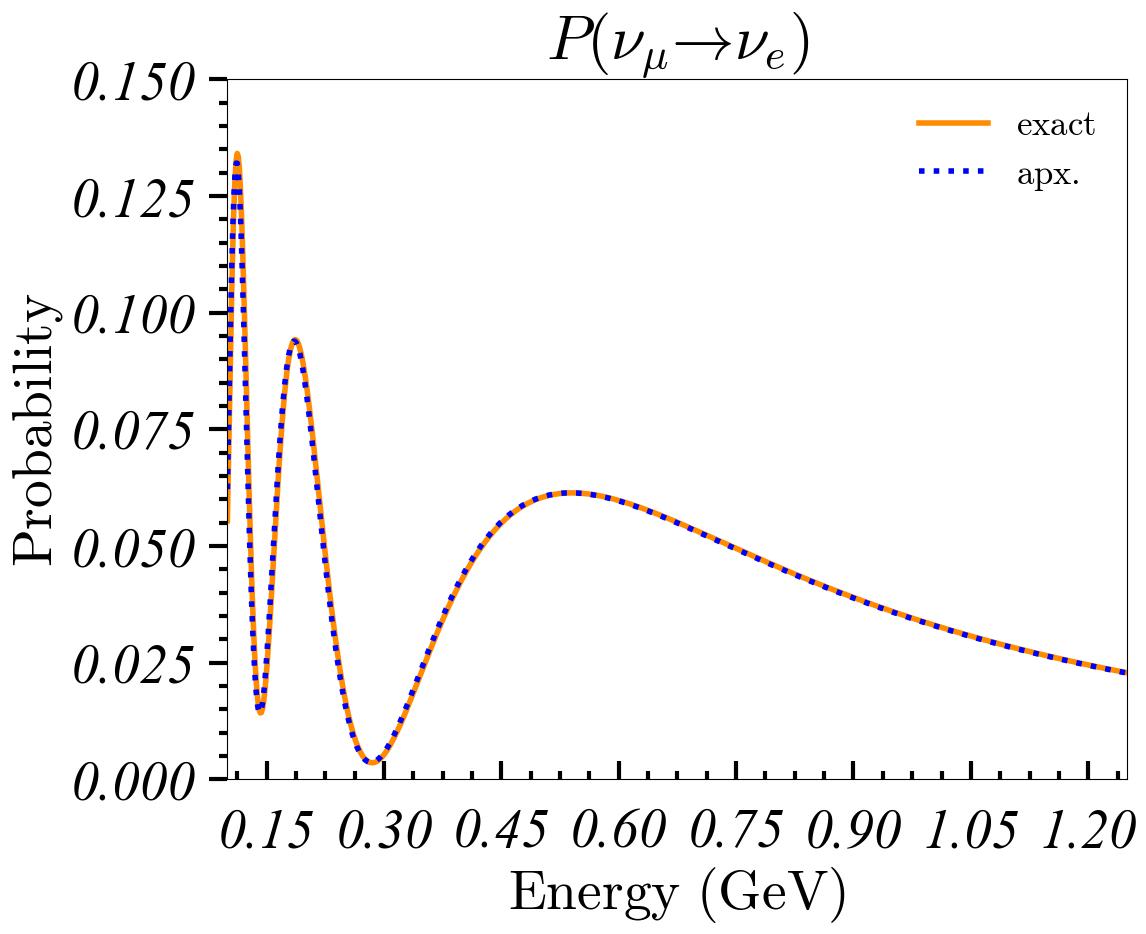}
    \caption{The orange solid line is the energy dependence of the oscillation probability and the dotted blue line is the approximated one in Eq.~\eqref{eq:prob-apx}.}
    \label{fig:prob-exact-apx}
\end{figure}

\begin{figure}[H]
    \centering
    \includegraphics[width=15cm]{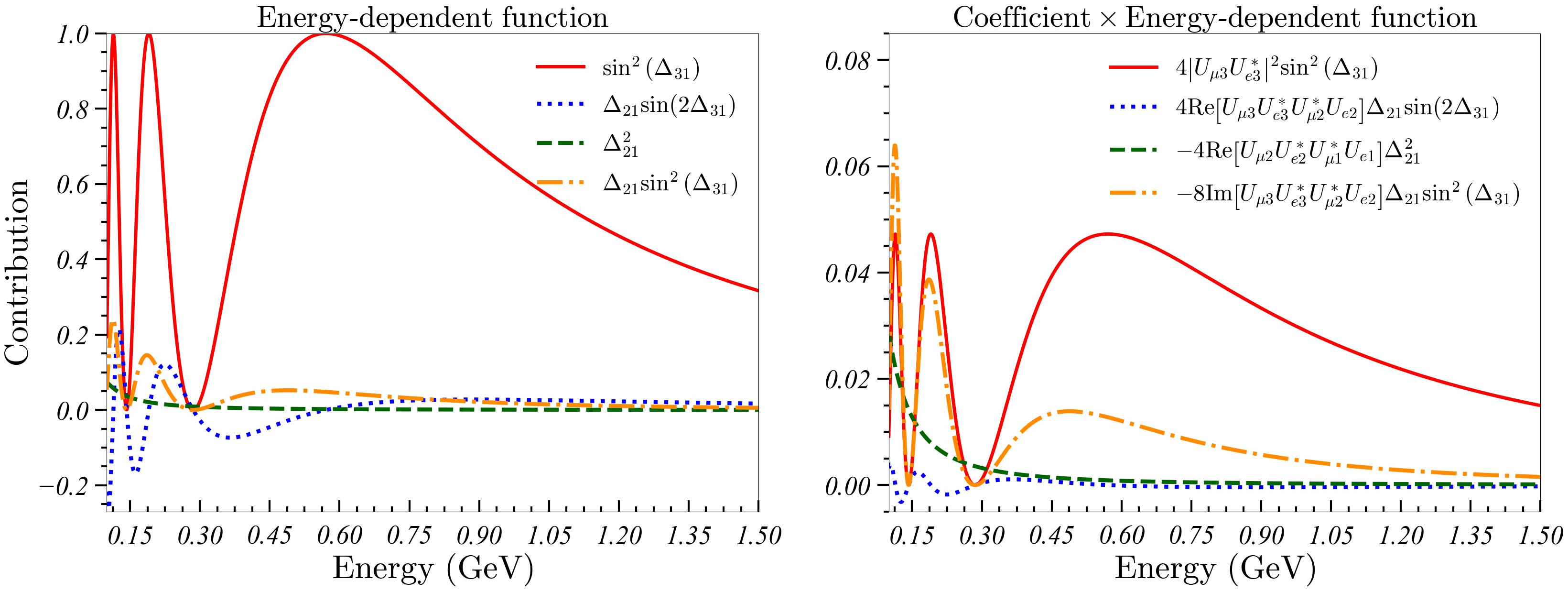}
    \caption{The contribution of each energy-dependent function to the oscillation probability. Figure on the left shows each of the energy-dependent functions. Figure on the right shows the contributions of each function with the coefficients multiplied. The solid red line, dotted blue line, dashed green line, and dash-dot orange line represent $\sin^2(\Delta_{31})$, $\Delta_{21}\sin(2\Delta_{31})$, $\Delta_{21}^2$, and $\Delta_{21}\sin^2(\Delta_{31})$, or the ones with the coefficients applied, respectively.}
    \label{fig:energy-dependent}
\end{figure}

Following Ref.~\cite{Sato:2000wv}, we take $V=U_{\rm PMNS}$ and retain terms 
up to the quadratic order in $\mathit{\Delta}m_{21}^2/\mathit{\Delta}m_{31}^2$ and $U_{e3}$.
The oscillation probability of the $\nu_{\mu \to e}$ appearance is given by
\begin{align}
    P^{\nu_{\mu\to e}} 
    \sim &\ 4 |U_{\mu 3} U_{e 3}^{*}|^2
    \sin^2(\Delta_{31}) \notag 
    \\  
    &+4 \mathrm{Re}\qty[U_{\mu 3} U_{e 3}^{*} U_{\mu 2}^{*} U_{e 2}]
    \Delta_{21}\sin(2\Delta_{31}) \notag 
    \\
    &-4\mathrm{Re}\qty[U_{\mu 2} U_{e 2}^{*} U_{\mu 1}^{*} U_{e 1}] 
    \Delta_{21}^2 \notag 
    \\ 
    &-8 \mathrm{Im}\qty[U_{\mu 3} U_{e 3}^{*} U_{\mu 2}^{*} U_{e 2}]
    \Delta_{21}\sin^2\qty(\Delta_{31})
    \label{eq:prob-apx}
\end{align}
where $\Delta_{jk}\equiv \mathit{\Delta}m^2_{jk}L/4E$.
The same formula is true for $\bar \nu_{e \to \mu}$ and the sign of $\mathrm{Im}\qty[U_{\mu 3} U_{e 3}^{*} U_{\mu 2}^{*} U_{e 2}]$ is flipped
for $\bar \nu_{\mu \to e}$ and $\nu_{e \to \mu}$.
One can see the quality of the approximated formula in Eq.~\eqref{eq:prob-apx} in Fig.~\ref{fig:prob-exact-apx}.
The energy dependence of the total oscillation probability is shown as the orange solid line
whereas the sum 
of four lines in the right panel of Fig.~\ref{fig:energy-dependent} is the blue dotted line, where the difference is invisible.
In Fig.~\ref{fig:energy-dependent}, we show the contribution of each energy-dependent function to the oscillation probability. 
We take the following reference values for three-generation model, which correspond to the arithmetic average of the best fit points in \cite{ParticleDataGroup:2022pth} from three groups, except for $\delta_{\rm CP}$. 
In this paper, we consider only the normal mass ordering. Unless otherwise stated, the three-generation oscillation probabilities are calculated using these parameters.
In the left panel, we show each of the energy-dependent functions. we show in the right panel the contributions of each function with the coefficients multiplied. The solid red line, dotted blue line, dashed green line, and dash-dot orange line correspond to $\sin^2(\Delta_{31})$, $\Delta_{21}\sin(2\Delta_{31})$, $\Delta_{21}^2$, and $\Delta_{21}\sin^2(\Delta_{31})$, or the ones with the coefficients applied, respectively. 
As can be seen from Eq.~\eqref{eq:prob-apx}, the contribution of $\sin^2(\Delta_{31})$ is generally large in all energy regions, but it is particularly prominent in the relatively high-energy region. The contributions of the other energy-dependent functions become larger at lower energies. Therefore, in order to fit the second, third, and fourth terms in Eq.~\eqref{eq:prob-apx}, it is important to observe the energy dependence at lower energy regions.
\begin{table}[H]
    \centering
    \begin{tabular}{c|c|c|c|c|c}
         $\theta_{12}$
         &$\theta_{13}$
         &$\theta_{23}$
         &$\delta_{\rm CP}$
         &$\mathit{\Delta} m_{21}^2/10^{-5}\ (\mathrm{eV}^2)$ 
         &$\mathit{\Delta} m_{31}^2/10^{-3}\ (\mathrm{eV}^2)$ 
        \\
         \hline 
         $\ang{33.9}$ & $\ang{8.49}$ & $\ang{48.1}$ & $\ang{270}$ & $7.43$ & $2.432$ \\
    \end{tabular}
    \caption{The reference values of three-generation oscillation parameters for the normal mass ordering.}
    \label{tab1}
\end{table}

In Eq.~\eqref{eq:prob-apx}, the oscillation probability can be expressed as a linear combination of four coefficients $C_{1-4}$ and known energy-dependent functions, as follows.
\begin{align}
    P^{\nu_{\mu\to e}} 
    =& C_1 \cdot \sin^2(\Delta_{31}) \notag
    \\
    &+C_2 \cdot \Delta_{21}\sin(2\Delta_{31}) \notag
    \\
    &+C_3 \cdot \Delta_{21}^2 \notag
    \\
    &+C_4 \cdot \Delta_{21}\sin^2\qty(\Delta_{31})\ .
    \label{eq:prob:coeff}
\end{align}

Therefore, we use this formula in following way.
The long-baseline experiments will
measure the 
appearance probability
such as $\nu_{\mu \to e}$
as a function
of energy.
We fit such histograms
by Eq.~\eqref{eq:prob:coeff}
with $C_{1-4}$ as parameters.
Note that, in principle, these four quantities are completely independent and can be measured independently.
They can be expressed in terms of the mixing matrix $U$ as follows.
%
\begin{align}
    C_1 &= 4 |U_{\mu 3} U_{e 3}^{*}|^2
    \label{eq:c1}
    \\
    C_2 &= 4 \mathrm{Re}\qty[U_{\mu 3} U_{e 3}^{*} U_{\mu 2}^{*} U_{e 2}]
    \\
    C_3 &= -4\mathrm{Re}\qty[U_{\mu 2} U_{e 2}^{*} U_{\mu 1}^{*} U_{e 1}]
    \\
    C_4 &= -8 \mathrm{Im}\qty[U_{\mu 3} U_{e 3}^{*} U_{\mu 2}^{*} U_{e 2}]\ .
    \label{eq:c4}
\end{align}
We now introduce a quantity $\xi$
\begin{align}
    \xi \equiv C_1(C_3-C_2) - C_2^2-\frac{C_4^2}{4}\ ,
\end{align}
which vanishes under the assumption that $U$ is unitary.
In general models, the energy dependence 
of the oscillation probability is not guaranteed to
be described by Eq.~\eqref{eq:prob:coeff}.
The best fit values of $C_{1-4}$ would not correspond
to the combination of the matrix elements in Eqs.~\eqref{eq:c1}-\eqref{eq:c4}
in such cases.
Since the ralation $\xi=0$ holds as a consequence of the unitarity of the $3\times3$ PMNS matrix, the $\xi$ parameter
evaluated by the best fit values
can be non zero,
for example, 
in the presence 
of the fourth generation
of neutrinos.
Thus, the measurement
of the $\xi$ parameter
provides us with
a test of unitarity
in the similar sense
as the CKM unitarity.
We demonstrate 
a unitarity test using the $\xi$ in the following 
sections.
%
\section{Statistical Analysis}
\label{sec3}
We here present the analyses of measuring the $\xi$ parameter in three and four generation
models, and discuss how well one can distinguish those cases.
Note that our analysis is not totally realistic as we ignore the matter 
effects. The study below should be interpreted as the first trial
of the unitarity test by assuming that the essential part of physics
is captured by this method.
\subsection{The least squares method}
\label{sec:3.1}
As shown in the previous section, the oscillation probability $P(E)$ at a given energy $E$ can be expressed as a linear combination, $\sum_k C_k B_k(E)$, where $B_k$ and $C_k$ are respectively the $k$-th energy-dependent function and its coefficient, such as
\begin{align}
    B_1(E)&=\sin^2(\Delta_{31})\ ,\ B_2(E)=\Delta_{21}\sin(2\Delta_{31})
    \notag \\
    B_3(E)&=\Delta_{21}^2\ ,\ B_4(E)=\Delta_{21}\sin^2(\Delta_{31})\ .
\end{align}
In general, the oscillation probability is expressed as follows,
\begin{align}
	P(E_j)=\sum_{k=1}^{m} C_k \cdot B_k(E_j)\ .
    \label{eq:osc-prob}
\end{align}
The discrete index $j = 1, \cdots, n$
represents the energy bins.
We express Eq.~\eqref{eq:osc-prob}
as a matrix multiplication
to a vector as:
\begin{align}
   \vb*P =\mathscr{B}\vb*{C}, 
   \label{eq:3gen-model}
\end{align}
where
\begin{align}
  \vb*{P} &\equiv \mqty(P(E_1) 
                      \\
                      P(E_2)
                      \\
                      \vdots
                      \\
                      P(E_n))
                      ,\ \ 
                      \vb*{C}\equiv\mqty(C_1 
                        \\
                        \vdots
                        \\
                        C_m),\ \ 
                        \vb*{B}_k \equiv \mqty(B_k(E_1)
                                        \\
                                         B_k(E_2)
                                        \\
                                         \vdots
                                        \\
                                         B_k(E_n))\ ,
\end{align}
and
\begin{align}
  \mathscr{B} &\equiv \mqty(\vb*{B}_1 & \cdots & \vb*{B}_m)\ .
\end{align}
When we perform the coefficient fit, the observed oscillation probability $P_j^{\rm obs}$ at a given energy bin is fitted using Eq.~\eqref{eq:3gen-model}. 
The $\chi^2$ value can be defined as
Using the observed number of events $N^{\mathrm{obs}}_{\mathrm{far}}$ and the expected number of events $N^{\mathrm{exp}}_{\mathrm{far}}$ at the far detector, the $\chi^2$ value can be defined as
\begin{align}
    \chi^2 &\equiv \sum_{j=1}^{n} \frac{\qty(N^{\mathrm{obs}}_{\mathrm{far},\ j}-N^{\mathrm{exp}}_{\mathrm{far},\ j})^2}{\qty(\mathit{\Delta}N^{\mathrm{obs}}_j)^2}
    \notag
    \\
    &=\sum_{j=1}^{n} \frac{\qty(N^{\mathrm{obs}}_{\mathrm{far},\ j}-\tilde{N}^{\mathrm{obs}}_{\mathrm{near},\ j}\cdot P(E_j))^2}
    {\qty(\mathit{\Delta}N^{\mathrm{obs}}_{\mathrm{far},\ j})^2
    +\qty(\mathit{\Delta}\tilde{N}^{\mathrm{obs}}_{\mathrm{near},\ j}\cdot P^{\mathrm{obs}}_j)^2}
    \notag
    \\
    &=\sum_{j=1}^{n} \frac{\qty(\qty(N^{\rm obs}_{{\rm far},\ j}/\tilde{N}^{\rm obs}_{{\rm near},\ j})-( \mathscr{B}C)_j)^2}{\qty(P^{\rm obs}_j)^2
    \cdot \qty[\qty(\frac{{\it \Delta}N^{\rm obs}_{{\rm far},\ j}}{N^{\rm obs}_{{\rm far},\ j}})^2 
    +\qty(\frac{{\it \Delta}N^{\rm obs}_{{\rm near},\ j}}{N^{\rm obs}_{{\rm near},\ j}})^2]}
    \notag
    \\
    &=\sum_{j=1}^{n} \frac{\qty(P_j^{\rm obs} - ( \mathscr{B}C)_j)^2}{\qty({\it \Delta}P_j^{\rm obs})^2}
    \notag
    \\
    &=\qty(\vb*{P}^{\rm obs}-\mathscr{B}\vb*{C})^{\sf T}W\qty(\vb*{P}^{\rm obs}-\mathscr{B}\vb*{C})\ ,
    \label{eq:leastchi}
\end{align}
where
\begin{align}
    \vb*{P}^{\rm obs} \equiv \mqty(P_1^{\rm obs}
                      \\
                      P_2^{\rm obs}
                      \\
                      \vdots
                      \\
                      P_n^{\rm obs}),\ \ 
    W\equiv {\rm diag}\qty(\frac{1}{\qty({\it \Delta}P_1^{\rm obs})^2}\ ,\cdots,\ \frac{1}{\qty({\it \Delta}P_{n}^{\rm obs})^2})\ .
\end{align}
Using $\tilde{N}^{\rm obs}_{\rm near}$, which is corrected so that $N^{\rm obs}_{\rm far}/\tilde{N}^{\rm obs}_{\rm near}$ gives the probabilities, the $N^{\rm exp}_{\rm far}$ can be written as $\tilde{N}^{\rm obs}_{\rm near}\cdot P(E)$. $\tilde{N}^{\rm obs}_{\rm near}$ different from $N^{\rm obs}_{\rm near}$. 
The ${\it \Delta} P_j^{\rm obs}$ parameter represents the 
statistical error of the observed oscillation probability
in each energy bin, $j$.
By using the error of the event numbers
measured at the near and the far detectors, ${\it \Delta} N_{{\rm far/near},\ j}^{\rm obs}$, in each bin,
${\it \Delta}P_j^{\rm obs}$ is given by 
\begin{align}
    \qty(\frac{\it{\Delta}P_j^{\rm obs}}{P_j^{\rm obs}})^2 &= \qty(\frac{\it{\Delta}N^{\rm obs}_{{\rm far},\ j}}{N^{\rm obs}_{{\rm far},\ j}})^2 + \qty(\frac{{\it \Delta}N^{\rm obs}_{{\rm near},\ j}}{N^{\rm obs}_{{\rm near},\ j}})^2
    \notag
    \\
    &\sim \qty(\frac{\it{\Delta}N^{\rm obs}_{{\rm far},\ j}}{N^{\rm obs}_{{\rm far},\ j}})^2\ .
    \label{eq:error}
\end{align}
Since $N^{\rm near} \gg N^{\rm far}$, the contribution to the statistical error is dominated by the one from far detectors.

The $C_k$ parameters to minimize $\chi^2$ is given by
\begin{align}
    \vb*{C}&=\qty(\mathscr{B}^{\sf T}W\mathscr{B})^{-1}\mathscr{B}^{\sf T}W\vb*{P}^{\rm obs}\ .
\end{align}
In more general cases where we combine 
multiple measurements from different beams  (represented as an index $l$ below such as $l = \{\nu_{\mu \to e}, \, \bar \nu_{\mu \to e} \}$),
we define $\chi^2$ as,
\begin{align}
    \chi^2 &\equiv \sum_l \sum_{j_l} \frac{\qty(P_{j_l}^{l,\rm obs} - ( \mathscr{B}C)^l_{j_l})^2}{\qty(\mathit{\Delta}P_{j_l}^{l,\rm obs})^2}
    \notag
    \\
    &=\sum_l \qty(\vb*{P}^{l,\rm obs}-\mathscr{B}^l\vb*{C}^l)^{\sf T}W^l\qty(\vb*{P}^{l,\rm obs}-\mathscr{B}^l\vb*{C}^l)
    \notag
    \\
    &=\mqty(\vdots \\ \vb*{P}^{l,\rm obs}-\mathscr{B}^l\vb*{C}^l \\ \vdots)^{\sf T}
    \mathrm{diag}(\cdots,W^l,\cdots)
    \mqty(\vdots \\ \vb*{P}^{l,\rm obs}-\mathscr{B}^l\vb*{C}^l \\ \vdots)\ .
\end{align}
For example, if there are two channels corresponding to CP (or T) conjugated modes, \\
i.e. $(\nu_{\mu\to e})+(\bar{\nu}_{\mu\to e})$ (or $(\nu_{\mu\to e})+(\nu_{e\to\mu})$), the expression takes the following form.
\begin{align}
	\chi^2 &\equiv \sum_{j=1}^{n} \qty[\frac{\qty(P_j^{\nu_{\alpha\to\beta}, \rm obs} - ( \mathscr{B}C)_j)^2}{\qty(\mathit{\Delta}P_j^{\nu_{\alpha\to\beta} ,\rm obs})^2}
	+\frac{\qty(P_j^{\bar{\nu}_{\alpha\to\beta} ,\rm obs} - ( \bar{\mathscr{B}}\bar{C})_j)^2}{\qty(\mathit{\Delta}P_j^{\bar{\nu}_{\alpha\to\beta} ,\rm obs})^2}]
    \notag
	\\
	&=\mqty(\vb*{P}^{\rm obs}-\mathscr{B}\vb*{C}\\
    \bar{\vb*{P}}^{\rm obs}-\bar{\mathscr{B}}\bar{\vb*{C}})^{\sf T}
    \mathrm{diag}(W,\bar{W})
    \mqty(\vb*{P}^{\rm obs}-\mathscr{B}\vb*{C}\\
    \bar{\vb*{P}}^{\rm obs}-\bar{\mathscr{B}}\bar{\vb*{C}})\ ,
\end{align}
Here, we denote quantities related to anti-neutrinos (or the T conjugate mode) 
as $\bar X$.
In this case, since the two oscillation probabilities have the same energy dependence and have identical coefficients except for the signs of each coefficient, we can consider $\mathscr{B}=\bar{\mathscr{B}}$, $C_k=\bar{C}_k,\ (1\le k \le m')$, and $C_k=-\bar{C}_k,\ (m'+1\le k \le m)$. Therefore,
\begin{align}
    \chi^2&=\mqty(\vb*{P}^{\rm obs}-\mathscr{B}\vb*{C}\\
    \bar{\vb*{P}}^{\rm obs}-S\mathscr{B}\vb*{C})^{\sf T}
    \mathrm{diag}(W,\bar{W})
    \mqty(\vb*{P}^{\rm obs}-\mathscr{B}\vb*{C}\\
    \bar{\vb*{P}}^{\rm obs}-S\mathscr{B}\vb*{C})
    \\
    S&\equiv \mqty(\vb*{1}_{m'} & 0 \\ 0 & -\vb*{1}_{m-m'})\ .
\end{align}
The expression can be further simplified as follows.
\begin{align}
    \chi^2&=\qty(\tilde{\vb*{P}}^{\rm obs}-\tilde{B}\vb*{C})^{\sf T}
    \tilde{W}
    \qty(\tilde{\vb*{P}}^{\rm obs}-\tilde{B}\vb*{C})
    \\
    \tilde{\vb*{P}}^{\rm obs} &\equiv \mqty(\vb*{P}^{\rm obs} \\ \bar{\vb*{P}}^{\rm obs})\ ,
    \ B\equiv \mqty(\mathscr{B}\\ S\mathscr{B})\ ,
    \ \tilde{W}\equiv \mathrm{diag}(W,\bar{W})\ .
\end{align}
%

\subsection{Number of Events at Hyper-Kamiokande}
\label{sec:3.2}
\begin{figure}[H]
    \centering
    \includegraphics[width=15cm]{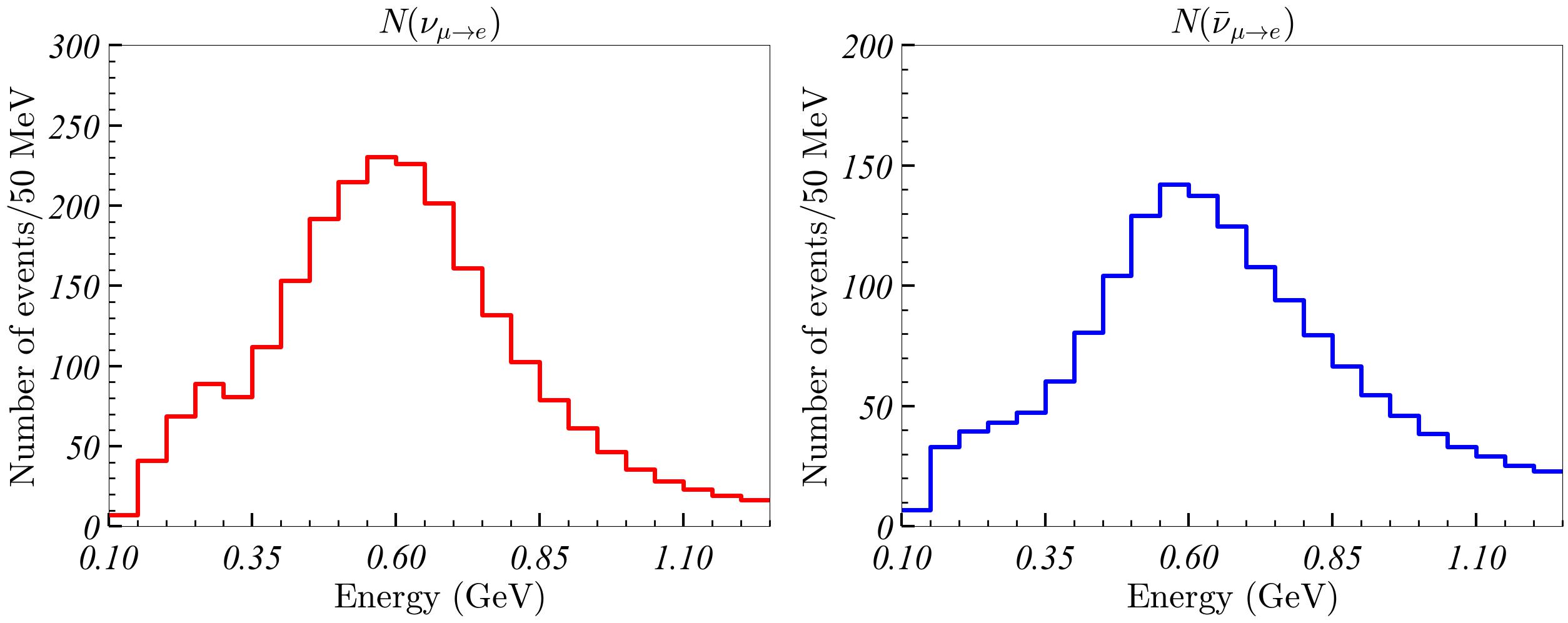}
    \caption{Number of events in the T2HK experiment. The left and right figures show the fluxes of $\nu_{\mu \to e}$ and $\bar{\nu}_{\mu \to e}$ at the far detector, respectively. Here, the CP phase is assumed to be $\delta_{\rm CP}=\ang{270}$. These expected number of events are calculated based on \cite{Hyper-Kamiokande:2018ofw}.}
    \label{fig:flux-T2HK}
\end{figure}

The $\nu_\mu$ and $\bar \nu_\mu$ beams are available
from the pion decays in the T2HK experiment~\cite{Hyper-Kamiokande:2018ofw}. The off-axis beam gives neutrinos to have energy peaked at $0.6~\rm GeV$. 
Considering the neutrino oscillation and the charged current quasielastic interactions
for the water Cherenkov detector, the expected number of events of the $\nu_{\mu \to e}$ and $\bar \nu_{\mu \to e}$ modes are obtained as in Fig.~\ref{fig:flux-T2HK}. The figures are
obtained based on the detector parameters given in Ref.~\cite{Hyper-Kamiokande:2018ofw} and the 
quasielastic neutrino cross sections in Ref.~\cite{Formaggio:2012cpf}.
The CP phase is chosen to be $\delta_{\rm CP} = \ang{270}$, and 
we apply a smearing of the neutrino spectrum based on a representative detector resolution of $\mathit{\Delta} E = 50 \ \mathrm{MeV}$~\cite{Hayato:2021heg}.

\begin{figure}[H]
    \centering
    \includegraphics[width=15cm]{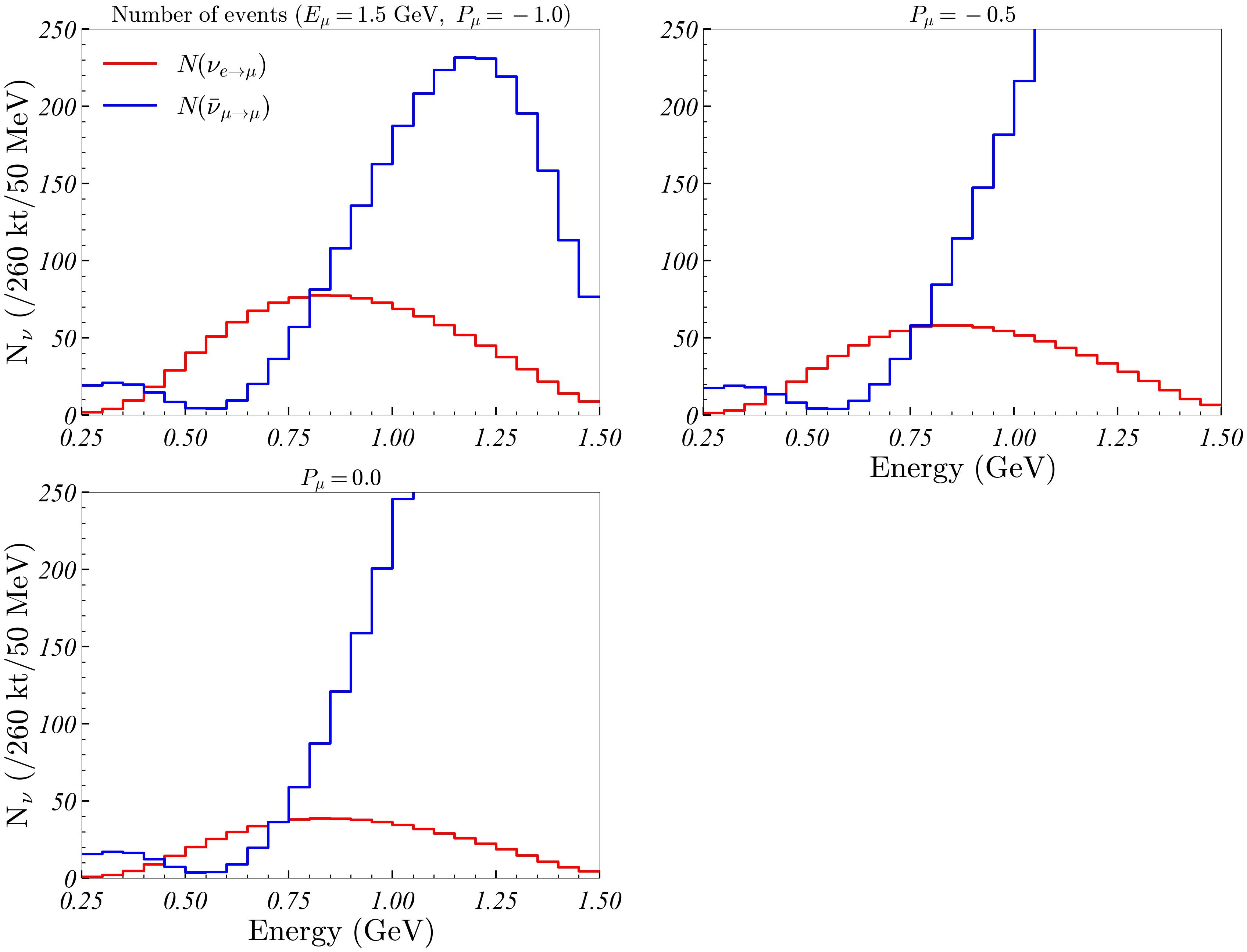}
    \caption{Neutrino flux measured at the Hyper-Kamiokande detector. The anti-muon beam energy is set to $1.5~\mathrm{GeV}$, and calculations are performed for four polarizations ($P_\mu=-1.0,\ -0.5,\ 0.0$). Here we set $\delta_{\rm CP}=\ang{270}$. The total number of muons is set to $10^{22}$.}
    \label{fig:flux-NF}
\end{figure}

The electron neutrino flux from the muon decays at the neutrino factory~\cite{Hamada:2022mua, Kitano:2024kdv} can be estimated by the
neutrino-energy ($E_\nu$) distribution of the (polarized) $\mu^+$
decay with a fixed energy $E_\mu$~\cite{Geer:1997iz,Barger:1999fs}. To obtain the event rate in each energy bin, one convolves the neutrino energy distribution with both the oscillation probability and the cross section for neutrino-nucleon scattering. The number of events is given by
\begin{align}
    N_j^{\nu_{e \to \mu}} &= \int_{E_j}^{E_{j+1}} \frac{dE_\nu}{E_\mu} \times \frac{12N_\mu\cdot V_d\cdot n_N}{\pi L^2} \times \gamma^2 \qty(\frac{E_\nu}{E_\mu})^2 \times \qty[\qty(1-\frac{E_\nu}{E_\mu}) -P_\mu\qty(1-\frac{E_\nu}{E_\mu})] \notag \\
    & \times P^{\nu_{e\to\mu}}(E_\nu) \times \sigma_{\nu_\mu}(E_\nu)\ ,
    \label{eq:flux_m2e}
    \\
    N_j^{\bar{\nu}_{\mu \to \mu}} &= \int_{E_j}^{E_{j+1}} \frac{dE_\nu}{E_\mu} \times \frac{2N_\mu\cdot V_d\cdot n_N}{\pi L^2} \times \gamma^2 \qty(\frac{E_\nu}{E_\mu})^2 \times \qty[\qty(3-\frac{2E_\nu}{E_\mu}) -P_\mu\qty(1-\frac{2E_\nu}{E_\mu})] \notag \\
    & \times P^{\bar{\nu}_{\mu \to \mu}}(E_\nu) \times \sigma_{\bar{\nu}_\mu}(E_\nu)\ ,
    \label{eq:flux_mb2mb}
\end{align}
where $V_d$ is the detector volume, $n_N$ is the number density of the nucleon in water,
and $L$ is the baseline length. The boost factor $\gamma$ is
that for the muon, i.e., $\gamma = E_\mu / m_\mu$ with $m_\mu$ the
muon mass.
We examine the polarization of the anti-muon beam, denoted by $P_\mu$, which is crucial for enhancing the $\nu_e$ flux in the forward direction.
As we plan to perform combined analyses with the T2HK experiment, the far detector is assumed to be Hyper-Kamiokande, and the baseline length is set to $L = 295 \ \mathrm{km}$.
In Fig.~\ref{fig:flux-NF}, we present the number of events $N^{\nu_{e \to \mu}}$ (represented by the red solid line) for the case of $\delta_{\rm CP} = \ang{270}$, with anti-muon polarization values $P_\mu = -1.0,\ -0.5,\ 0.0$.
The events are smeared by the detector resolution, $\mathit{\Delta} E = 50 \ \mathrm{MeV}$~\cite{Hayato:2021heg}, as in the T2HK analysis.
Additionally, we overlay the number of background events, $N_j^{\bar \nu_{\mu \to \mu}}$ (represented by the blue solid line). The total number of muons (which decay towards the Hyper-Kamiokande detector) is set to $10^{22}$ and set $E_\mu=1.5\ \rm GeV$.
%
%
%

\subsection{Background Subtraction and Statistical Uncertainties}
\label{sec:3.3}

The $\bar{\nu}_{\mu \to \mu}$ events shown in Fig.~\ref{fig:flux-NF} become background of the $\nu_{e \to \mu}$
channel if $\nu_{\mu}$ and $\bar \nu_\mu$ are indistinguishable at the Hyper-Kamiokande detector.
The separation of those events requires the charge identification of the muons
produced by the charged-current interactions, $\nu_\mu n\to \mu^- p$ and $\bar{\nu}_\mu p \to \mu^+ n$.
Although the neutron tagging of the final state can, in principle, distinguish the charges of muons~\cite{Beacom:2003nk,Huber:2008yx}, the efficiency is still under study~\cite{Beacom:2003nk,Super-Kamiokande:2023xup,Akutsu2019}.
Therefore, as in \cite{Kitano:2024kdv}, we account for the background from the charge misidentification
by assuming that the efficiency itself is already measured and known at the T2HK experiment.
Given the knowledge of the efficiency, one can simply subtract the expected number of background
by using the survival probability, $P(\bar \nu_{\mu \to \mu})$ measured at the T2HK
experiment.
We carry out the analysis under three different scenarios for the charge identification efficiency: $100\%$, $70\%$, and $0\%$, corresponding to optimistic, intermediate, and pessimistic cases, respectively.
This correction also modifies the statistical error of $P^{\nu_{e\to\mu}}$, as shown in \cite{Kitano:2024kdv}.
As we will see, the ability of charge identification is, in the end, not quite important in the analysis.
Even in the totally indistinguishable case, one can subtract the background statistically.
Also, the background events are suppressed near the energy region where
the oscillation is maximized.

\subsection{Unitarity Test}
\label{sec:3.4}
This analysis was carried out under the assumption of an experimental setup with $\nu_\mu$, $\bar{\nu}_\mu$, and $\nu_e$ beams, as expected in facilities like T2HK and a future neutrino factory with the $\nu_e$ beam at J-PARC.
We use energy range of $0.1 \sim 1.2\ \rm GeV$ for T2HK and $0.25\sim 1.45\ \rm GeV$ for neutrino factory. 
The difference of the low energy threshold stems from
that in the final states in the charged current
processes. It is $e^-$ or $e^+$ for T2HK and $\mu^-$ for the neutrino factory where
more energy is needed to produce muons.

We generate one million random virtual-experiments, and perform $\chi^2$ analysis to extract the coefficients. In generating the virtual-experiments, we consider the two scenarios: the standard three-generation model and a four-generation model with an $\rm eV$ scale sterile neutrino as a simple example for violation of $3\times3$ unitarity.
In the following, virtual-experiments generated under the three-generation assumption are denoted as ``three-generation events'', and those generated under the four-generation assumption are denoted as ``four-generation events''.
Then, we compare
the the best fit value of $\xi$
and its error 
in each case
to 
see the discrimination 
power of 
the unitarity test.
The statistical quantities used to draw the following figures are the mean and variance obtained from one million virtual-experiments.
For the four-generation model, as a sample point, we use the fit results provided in \cite{Dentler:2018sju,Parveen:2024bcc}, where the parameters $\theta_{12},\ \theta_{13},\ \theta_{23},\ \theta_{14},\ \theta_{24},\ \theta_{34},\ \delta_{\rm CP},\ \delta_{24},\ \delta_{34},\ \mathit{\Delta}m_{21}^2$, $\mathit{\Delta}m_{31}^2$, and $\mathit{\Delta}m_{41}^2$ are set to their best fit values.
The chosen values of $\theta_{14}$ and $\theta_{24}$ are not excluded by the latest results from MicroBooNE~\cite{MicroBooNE:2025nll}.
\begin{table}[H]
    \centering
    \begin{tabular}{c|c|c|c|c|c}
         $\theta_{12}$
         &$\theta_{13}$
         &$\theta_{23}$
         &$\theta_{14}$
         &$\theta_{24}$ 
         &$\theta_{34}$ 
        \\
         \hline 
         $\ang{34.3}$ & $\ang{8.53}$ & $\ang{49.3}$ & $\ang{5.7}$ & $\ang{5}$ & $\ang{20}$ 
         \\
         \multicolumn{6}{c}{}
         \\
         \multicolumn{2}{c|}{$\delta_{\mathrm{CP}}$} & \multicolumn{2}{c|}{$\delta_{24}$} & \multicolumn{2}{c}{$\delta_{34}$}
         \\
         \hline
         \multicolumn{2}{c|}{$-\ang{165.6}$} & \multicolumn{2}{c|}{$\ang{0}$} & \multicolumn{2}{c}{$\ang{0}$}
         \\ 
         \multicolumn{6}{c}{}
         \\
         \multicolumn{2}{c|}{$\mathit{\Delta} m_{21}^2/10^{-5}\ (\mathrm{eV}^2)$} 
         & \multicolumn{2}{c|}{$\mathit{\Delta} m_{31}^2/10^{-3}\ (\mathrm{eV}^2)$} 
         & \multicolumn{2}{c}{$\mathit{\Delta} m_{41}^2\ (\mathrm{eV}^2)$}
         \\
         \hline
         \multicolumn{2}{c|}{$7.5$} & \multicolumn{2}{c|}{$2.55$} & \multicolumn{2}{c}{$1$}
    \end{tabular}
    \caption{The reference values of four-generation oscillation parameters for the normal mass ordering.}
    \label{tab2}
\end{table}

\begin{figure}[H]
    \centering
    \includegraphics[width=16cm]{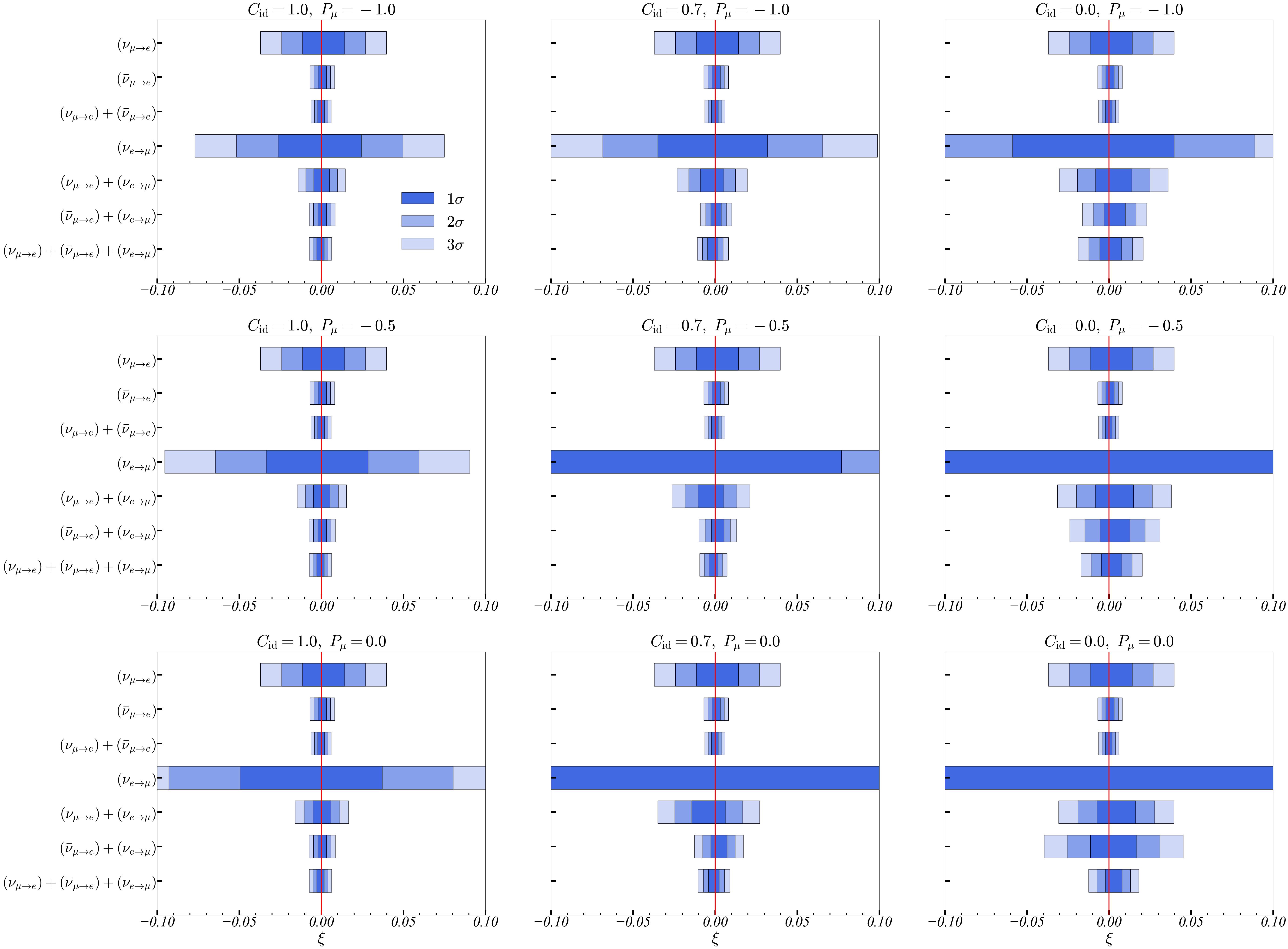}
    \caption{Unitarity test in terms of $\xi$. These figures show some comparisons of various channel combinations. These figures also compare different values of anti-muon beam polarization~$P_\mu$ and charge identification efficiency~$C_{\mathrm{id}}$. From top to bottom, the figures show the cases with $P_\mu=-1.0,\ -0.5,\ 0.0$. From left to right, the figures show the cases with $C_{\mathrm{id}}=1.0,\ 0.7,\ 0.0$. In these figures, the total number of muons is set to $10^{22}$. The blue regions represent the results of the fitting three-generation model to three-generation events, with the color intensity indicating the allowed region up to the $3\sigma$ level. The red solid vertical lines represent $\xi=0$.}
    \label{fig:UT-10-22-3gen}
\end{figure}

\begin{figure}[H]
    \centering
    \includegraphics[width=16cm]{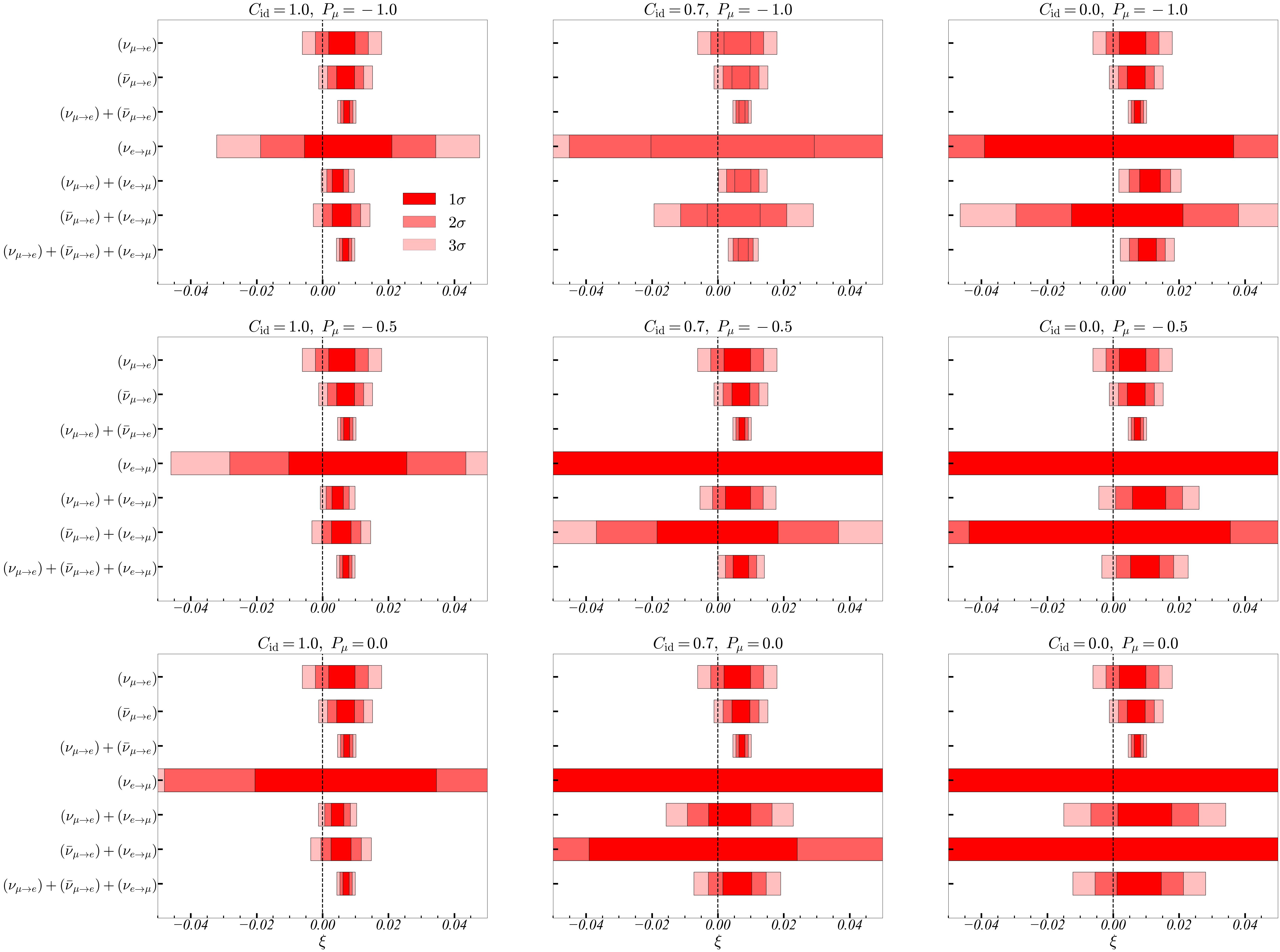}
    \caption{Unitarity test in terms of $\xi$. As in Fig~\ref{fig:UT-10-22-3gen}, these figures show some comparisons of various channel combinations and also compare different values of anti-muon beam polarization~$P_\mu$ and charge identification efficiency~$C_{\mathrm{id}}$. In these figures, the total number of muons is set to $10^{22}$. The red regions represent the results of the fitting three-generation model to four-generation events, with the color intensity indicating the allowed region up to the $3\sigma$ level. The black dashed vertical lines represent $\xi=0$.}
    \label{fig:UT-10-22-4gen}
\end{figure}

In this study, all analyses are performed by fitting with the model based on the three-generation theory, i.e., with Eq.~\eqref{eq:prob:coeff}.
We present the results of testing unitarity in Fig.~\ref{fig:UT-10-22-3gen} and Fig.~\ref{fig:UT-10-22-4gen}. In these figures, we show comparisons for the choices of different combinations of channels. We alse compare different values of anti-muon polarization~$P_\mu$ and charge identification efficiency~$C_{\rm id}$. We set the polarization to $P_\mu=-1.0,\ -0.5,\ 0.0$ (from top to bottom) and the charge identification efficiency to $C_{\mathrm{id}}=1.0,\ 0.7,\ 0.0$ (from right to left). The contrast in the color shading corresponds to the $1\sigma,\ 2\sigma,$ and $3\sigma$ allowed regions. The red solid (in Fig.~\ref{fig:UT-10-22-3gen}) or black dashed (in Fig.~\ref{fig:UT-10-22-4gen}) vertical lines represent $\xi=0$. In these figures, the total number of muons (which decay towards the Hyper-Kamiokande detector) in the neutrino factory is set to $10^{22}$.
Figure.~\ref{fig:UT-10-22-3gen} and Fig.~\ref{fig:UT-10-22-4gen} show the results of fitting the three-generation model to the three-generation events and to the four-generation events, respectively.

\begin{figure}[H]
    \centering
    \includegraphics[width=15cm]{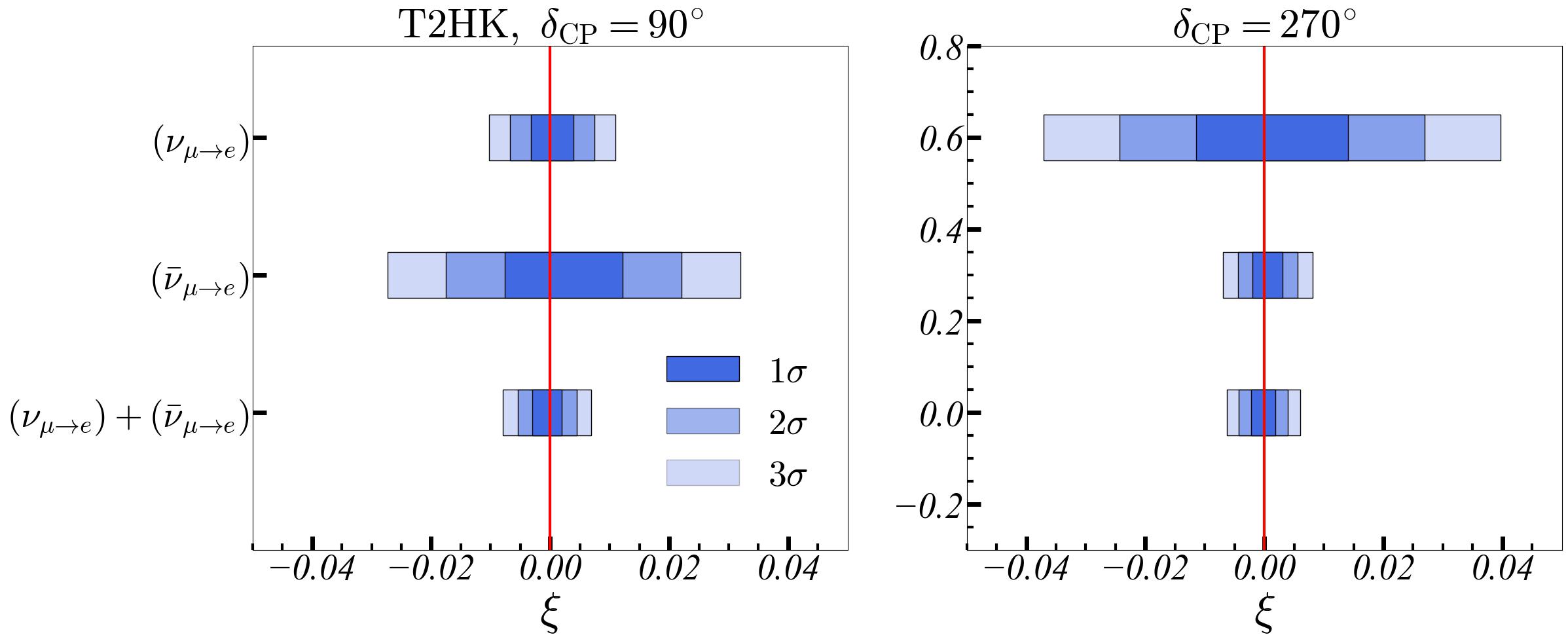}
    \caption{Unitarity test in terms of $\xi$. These figures show the results of the analysis using the channels in T2HK. These figures also compare different values of CP phase $\delta_{\rm CP}(=\ang{90},\ \ang{270})$. The blue regions represent the results of the fitting three-generation model to three-generation events, with the color intensity indicating the allowed region up to the $3\sigma$ level. The red solid vertical lines represent $\xi=0$.}
    \label{fig:UT-T2HK-3gen-vary-delta}
\end{figure}

Since Fig.~\ref{fig:UT-10-22-3gen} shows the results of fitting the three-generation model to the three-generation events, it indicates consistency with $\xi=0$ in the context of the testing unitarity.
All analyses show results consistent with $\xi=0$; however, the uncertainty varies depending on the observation channel or combination. Especially when using only the beam from the neutrino factory, i.e., only $(\nu_{e\to\mu})$, the uncertainty varies significantly depending on the polarization of the anti-muon beam and the charge identification efficiency. However, combining it with T2HK yields a better fit than a single-channel analysis, even in the worst-case scenario $(C_{\rm id}=0.0,\ P_\mu=0.0)$.
Another noteworthy point is that, in the T2HK analysis, the fit for $(\bar{\nu}_{\mu\to e})$ shows several times smaller uncertainty than that for $(\nu_{\mu\to e})$. 
In Fig.~\ref{fig:UT-T2HK-3gen-vary-delta}, we present the results of the unitarity test for two values of the CP phase 
$\delta_{\rm CP}$: $\ang{90}$ and $\ang{270}$. The results shown here are obtained from the analysis using the channels in T2HK. It is found that in the analysis using $\nu_{\mu\to e}$ and $\bar{\nu}_{\mu\to e}$, the uncertainties of $\xi$ depend on $\delta_{\rm CP}$.
As shown in Fig.~\ref{fig:UT-T2HK-3gen-vary-delta}, the opposite behavior of the uncertainty between $\delta_{\rm CP}=\ang{90}$ and $\delta_{\rm CP}=\ang{270}$ arises from the correspondence between the transformation $\delta\to -\delta$ and the CP-conjugation of the neutrino oscillation and anti-neutrino oscillation.
The oscillation probabilities of neutrinos and anti-neutrinos have the same energy dependence, but the CP-violating terms differ in the sign of their coefficients. As a result, the uncertainty of $\xi$ differs between neutrinos and antineutrinos. However, which of the two yields a smaller uncertainty depends on the value of $\delta_{\rm CP}$.

Fig.~\ref{fig:UT-10-22-4gen} shows the results of the fitting of the $\xi$ parameter with the 
four-generation events
based on Eq.~\eqref{eq:prob:coeff}.
We see inconsistencies with the unitarity condition, $\xi = 0$.
Since we are performing data fitting with a wrong model, 
the mean value of the best fit values can vary depending on the channel.
Although the differences in the $\xi$ parameter in each mode in Fig.~\ref{fig:UT-10-22-4gen} are mild,
the mean values of the best fit values of $\xi$ are not quite important and not physical. The important point is how far it is from $\xi=0$ in terms of the uncertainties.
It is found that an analysis using two appearance channels in T2HK, namely $(\nu_{\mu\to e})+(\bar{\nu}_{\mu\to e})$, is sufficient to test the violation of unitarity. In other words, $\xi=0$ is inconsistent at more than $3\sigma$.

On the contrary, in the analysis using only $(\nu_{e\to\mu})$ from the neutrino factory, the number of events is insufficient, making it difficult to test unitarity. In addition, in analyses using channels from the neutrino factory, the polarization of the anti-muon beam and the charge identification efficiency are important factors.
As in Fig.~\ref{fig:UT-10-22-3gen} and Fig.~\ref{fig:UT-10-22-4gen} also shows that combining multiple channels reduces the uncertainty.

In Fig.~\ref{fig:UT-10-22-3gen} and Fig.~\ref{fig:UT-10-22-4gen}, other than the statistical advantages, having a low energy threshold at T2HK helps
to have better sensitivities to the unitarity test compared to the neutrino factory.
As one can see from the approximate expression in Eq.~\eqref{eq:prob:coeff}, in the relatively high-energy region, 
$\sin^2(\Delta_{31})$ gives the dominant contribution, making it difficult to extract coefficients other than $C_1$. 
Therefore, access to the low energy region is important in order to extract the information of other coefficients which are necessary to conduct the unitarity
test.
In addition, by combining T2HK and the neutrino factory, CP-conjugate, T-conjugate, and CPT-conjugate channels can be tested independently. Since these channels are combinations of two appearance channels, they provide a more powerful test of unitarity than a fit using a single channel alone.

\begin{figure}[H]
    \centering
    \includegraphics[width=15cm]{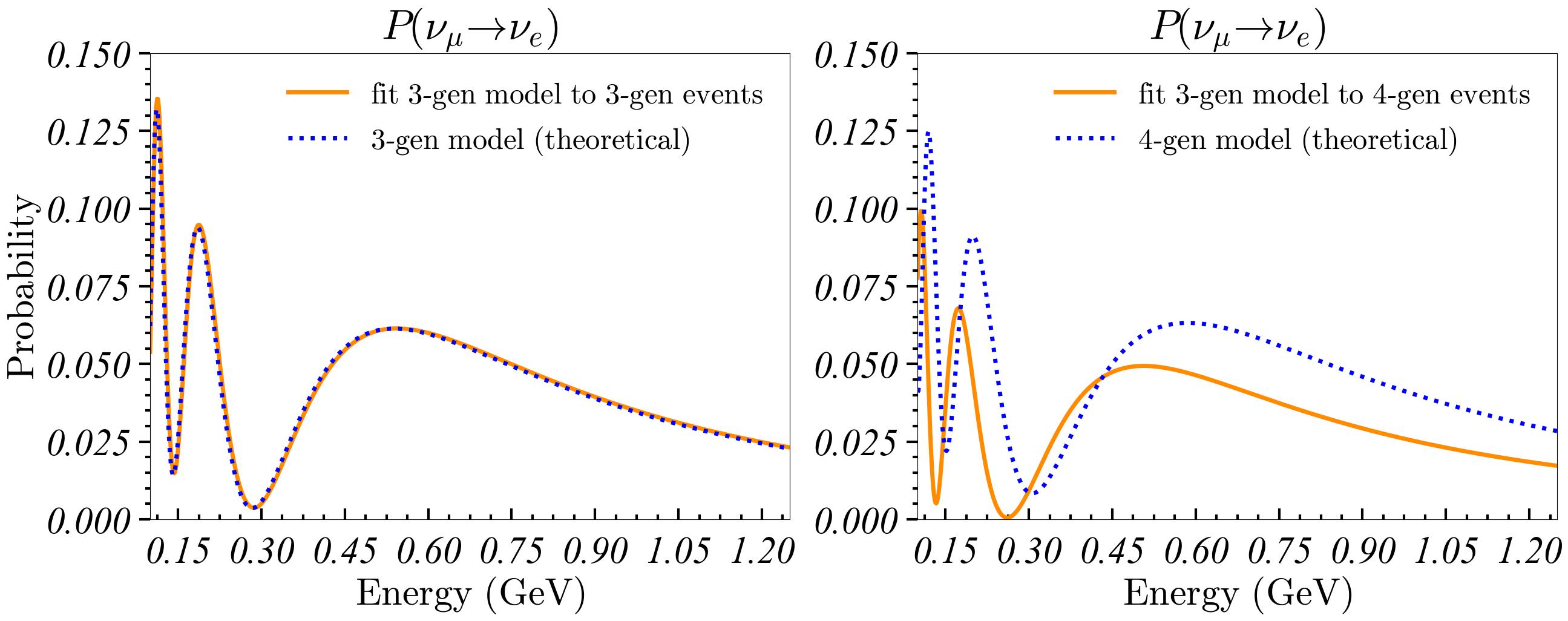}
    \caption{These figures show the oscillation probabilities in three-generation (left panel) and four-generation (right panel), respectively. The orange solid line is drawn using the results of the $\chi^2$ fit of the three-generation model to the three-generation events (left panel) and to the four-generation events (right panel). The blue dotted line represents the theoretical curve of the three-generation model (left panel) and that of the four-generation model (right panel).}
    \label{fig:prob-mean_value}
\end{figure}

Finally, we present the oscillation probabilities obtained using the coefficients fitted in Fig.~\ref{fig:prob-mean_value}. These plots are drawn on the results of the $\chi^2$ analysis using channels $(\nu_{\mu\to e})+(\bar{\nu}_{\mu\to e})$.
The left panel shows the oscillation probabilities in three-generation.
The orange solid line is drawn using the results of the $\chi^2$ fit of the three-generation model to the three-generation events, while the blue dotted line represents the theoretical curve obtained from the oscillation probability given in Eq.~\eqref{eq:prob:coeff} with the parameters listed in Table~\ref{tab1}. Although it should have an error bar, the curve is obtained using only the average of the coefficients from one million fits.
The right panel shows the oscillation probabilities in the four-generation case. The orange solid line is the curve drawn using the results of the $\chi^2$ fit of the three-generation model to the four-generation events, and the blue dotted line represents the theoretical curve of the four-generation model with the parameters listed in Table~\ref{tab2}.

As shown in the left panel of Fig.~\ref{fig:prob-mean_value}, the results of fitting the three-generation model to the three-generation events agree well with the theoretical curve of the three-generation model under ideal statistics. Although this is an expected result, it confirms that the coefficient fit has been performed consistently without any contradictions.
On the contrary, the right panel of Fig.~\ref{fig:prob-mean_value} shows that the results of fitting the three-generation model (Eq.~\eqref{eq:prob:coeff}) to the four-generation events differ from the theoretical curve of the four-generation model. This curve is also obtained under ideal statistics. Although this is also expected from the results of Fig.~\ref{fig:UT-10-22-4gen}, it shows that when the four-generation events are incorrectly fitted with the three-generation model, inconsistencies appear even at the level of the oscillation probability.


%
%

\section{Summary}
\label{sec4}
We propose a method for testing unitarity without assuming a specific parameterization of PMNS matrix, and demonstrate its applicability and sensitivity to future experiments.
In this paper, we conducte an analysis using appearance channels of neutrino oscillation, assuming future experiments such as T2HK~\cite{Hyper-Kamiokande:2018ofw} and the neutrino factory with a 
$\nu_e$ beam at J-PARC~\cite{Hamada:2022mua, Kitano:2024kdv}. In addition, this study assumes vacuum oscillations, both for simplicity and because the baseline in the considered experiments are relatively short.

By considering parameters that are the four-generation best fit points, we demonstrated that T2HK alone can provide sufficient sensitivity to test unitarity in terms of $\xi$. Moreover, combining multiple appearance channels provides even better sensitivity.
When combining T2HK and the neutrino factory, we can test the unitarity by using T-conjugate channel ($(\nu_{\mu\to e})+(\nu_{e\to \mu})$), CP-conjugate channel ($(\nu_{\mu\to e})+(\bar{\nu}_{\mu\to e})$), or CPT-conjugate channel ($(\bar{\nu}_{\mu\to e})+(\nu_{e\to\mu})$), independently
%

As we mentioned in Section~\ref{sec2}, when fitting the coefficients by observing the energy dependence, it is important to observe the relatively low-energy region. In this regard, the T2HK experiment, with final-state particles being $e^-$ or $e^+$, allows observations down to lower energies compared to the neutrino factory. Therefore, in addition to its statistical advantages, the low-energy threshold of T2HK helps improve the sensitivity.

The present analysis is performed in vacuum, and the overall sensitivity is expected to get worse when matter effects are taken into account. Especially, the channels related by CP-conjugation will be directly impacted by matter effects. On the other hand, the combination of T-conjugate channels may remain to give similar sensitivity as T-violation is insensitive to the matter effects.  
Although a detailed quantitative study is needed,
the current analysis using $(\nu_{\mu\to e})$ and $(\nu_{e\to\mu})$ in vacuum may catch
the overall feature of the whole analysis.
We leave the analysis including the matter effects and other systematic uncertainties for future work.

Unitarity test for lepton mixing is an important task for particle physics and 
must be done in future neutrino experiments. In this study, we demonstrated the possibility of testing unitarity in long-baseline neutrino oscillation experiments,
and found that meaningful independent measurements are possible at T2HK and future neutrino factories.

\section*{Acknowledgements}
This work was in part supported
by JST SPRING Japan Grant Number JPMJSP2178 (S.S.).
This work is also supported in part by JSPS KAKENHI Grant-in-Aid for
Scientific Research (No.~22K21350~[RK], No.~25H01524~[JS]) and
the U.S.-Japan Science and Technology Cooperation Program in High Energy Physics~(2025-20-2~[RK]).

\appendix
\section{Appendix}
\subsection{Virtual-experiment}
\label{appendix:A}

In generating virtual-experiments, since the observed number of events is an integer, we assume a binomial distribution $B(n,p)$ in this study.
In a binomial distribution, $n,\ p$ correspond to the number of non-oscillated events and the oscillation probability, respectively. 
Since values of $n$ and $p$ can be assigned to each energy bin, the virtual-experiments are randomly generated according to the binomial distribution in each bin.

\subsection{The treatment of zero-event bin}
\label{appendix:B}

When the expected number of events in a bin is small, it often happens It often happens that zero events are observed due to the binomial nature of the distribution.
Including zero-event bins in the analysis causes the estimation of $C_i$ by Eq.~\eqref{eq:leastchi} to break down.
To address this, one could simply exclude bins with zero events; nevertheless, in the present analysis, we account for the possibility of observing zero events within the statistical framework.
For this reason, we include this possibility in the analysis as a zero correction.
First, we define the zero threshold $p_0$ as the value at which the probability of observing zero events equals the probability of observing one or more events. That is, considering a random variable $X$ following a binomial distribution, we define the $p_0$ such that the probability of $X=0$, namely $P(X=0)$, equals $P(X\ge1)$,
\begin{align}
    P(X=0) &= P(X \ge 1) \notag
    \\
    \qty(1-p_0)^n &= 1 - \qty(1-p_0)^n \notag
    \\
    p_0 &= 1-\qty(\frac{1}{2})^{1/n}\ .
    \label{eq:zero-correction}
\end{align}
Thus, the expectation value corresponding to this probability $np_0$ (which asymptotes to $\log 2 \sim 0.7$ as $n \to \infty$) is added to the bin so that contributions to $\chi^2$ from 
those zero bins remain finite.
This treatment would not affect the fitting of $C_{1-4}$ when the total value 
of $\chi^2$ is large enough, {\it i.e.,} for large enough statistics. We will check
how this treatment affects the evaluation of the $\xi$ parameter
for various assumptions of neutrino fluxes in the following.

There are two methods for applying this zero correction, i.e., applying to all bins or only to those with zero events.
In the following, we demonstrate that two of the treatments do not show significant differences
in particular for the case with
large enough statistics.
Also, we compare the behavior of statistical uncertainties in these methods as we decrease statistics in order to understand
qualitative features.
We use the all-bin method in the analyses in this paper. Note, however, that the results are not significantly affected
by these treatments as we have enough statistics by assuming enough neutrino fluxes in each experiment.

\begin{figure}[H]
    \centering
    \includegraphics[width=15cm]{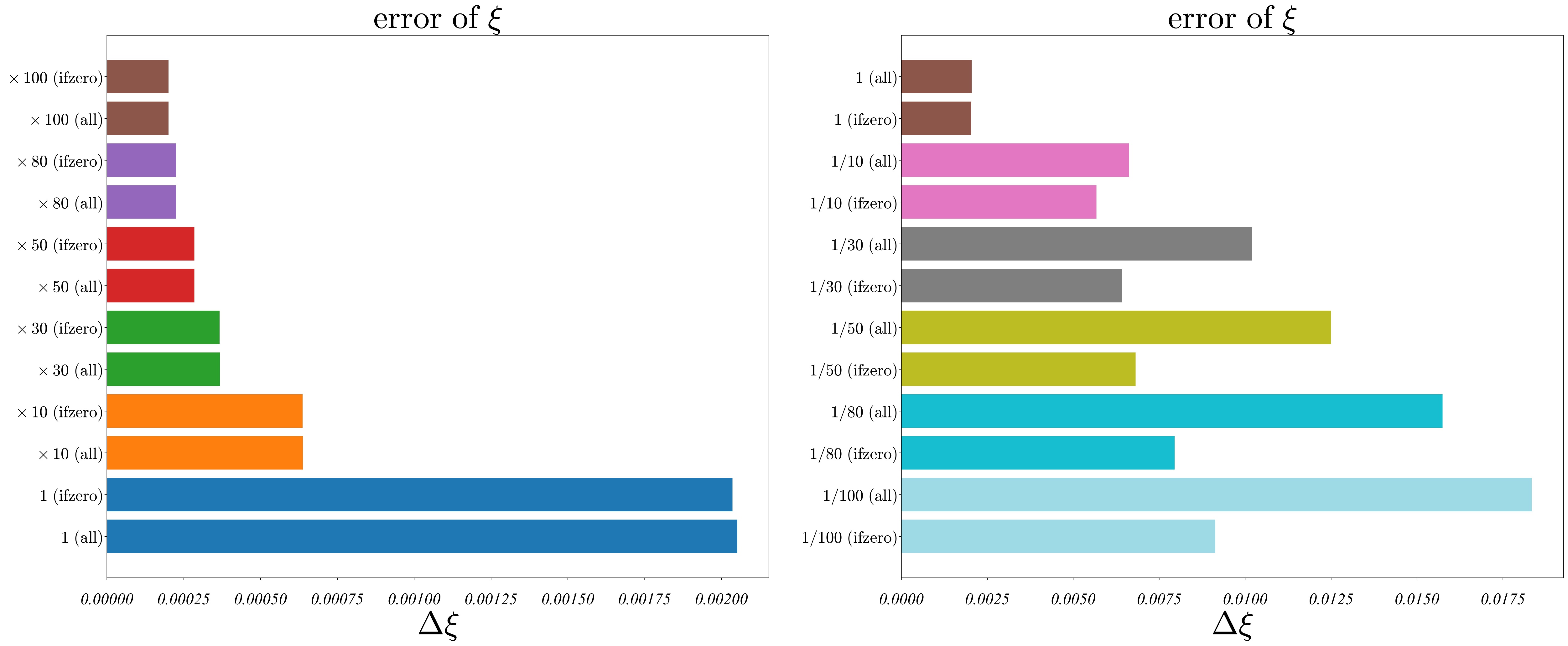}
    \caption{Effect of the zero correction on $\xi$. In this figures, we use $(\nu_{\mu \to e})$ and $(\bar{\nu}_{\mu \to e})$ from T2HK to extract the coefficients and the results are presented in terms of $\xi$. The left panel corresponds to the case where the number of events is increased. In contrast, the right panel shows the case where the number of events is decreased. The label ``all" corresponds to the analysis where the zero correction factor is added to all bins, while ``ifzero" corresponds to the analysis where the correction is applied only to bins that are observed as zero events when generating virtual-experiments.}
    \label{fig:zerocorrection}
\end{figure}

Figure~\ref{fig:zerocorrection} shows the contribution of the zero correction for different expected numbers of events.
In these figures, the analysis is performed using the combined channels $(\nu_{\mu\to e})+(\bar{\nu}_{\mu\to e})$ from T2HK. These figures also show the errors resulting from fitting the three-generation events with the three-generation model. The left panel corresponds to the case where the number of events is increased, while the right panel shows the case where the number of events is decreased. Since evaluating the goodness of fit critically depends on the magnitude of the error, only the error on $\xi$ is plotted.
Fig.~\ref{fig:zerocorrection} shows that when the expected number of events is sufficiently large, and the probability of observing zero events is negligible, the choice of zero-correction method has little effect on the resulting errors.
On the other hand, when bins with expected values around 1 or below $np_0$ are included, noticeable differences emerge depending on how the zero correction is applied, resulting in significant variations in the error.

\bibliographystyle{./utphys.bst}
\bibliography{./UnitarityCheck.bib}

\end{document}